# VTXO – VIRTUAL TELESCOPE FOR X-RAY OBSERVATIONS

Kyle Rankin,[*] Steven Stochaj,[†] Neerav Shah,[‡] John Krizmanic,[§]

and Asal Naseri[**]

The Virtual Telescope for X-Ray Observations (VTXO) is a conceptual mission under development to demonstrate a new instrument for astronomical observations in the X-ray band using a set of 6U CubeSats. VTXO will use a Phase Fresnel Lens, which has the potential to improve the imaging resolution several orders of magnitude over the current state-of-the-art X-ray telescopes. This technique requires long focal lengths (>100 m), which necessitates the lens and camera be on separate spacecraft, flying in precision formation. This work presents the results from a model developed to determine the ΔV requirements to maintain formation, for both solar and galactic X-ray observations, from a Geostationary Transfer Orbit.

**Introduction**

The VTXO (Virtual Telescope for X-Ray Observations) is a concept development mission for the next generation X-Ray telescope. This mission aims to fly a Phase Fresnel Lense which will provide several orders of magnitude better resolution then is possible with current state of the art X-ray optics. However, the Phase Fresnel Lense poses a significant challenge to spacecraft design, in that it has a focal length in excess of 100m. In order to meet these requirements, the VTXO mission will split the telescope components onto two separate vehicles. The first vehicle will carry the lense, while the second would carry the X-ray detector, these two vehicles would then fly in a formation approximating a rigid telescope.

The mission design for the VTXO spacecraft calls for the two vehicles to hold a rigid formation near apogee, during which time the two spacecraft will perform scientific observations for a short period of time (1h – 3h). While away from apogee the two vehicles will reposition themselves for the next iteration of the observations. The objective of this work was to understand the impact of the telescope pointing direction, different observation periods, changes in telescope focal length, and generally to characterize how the fuel consumption of the telescope varies with these parameters. In order to do this, a simulation was built based on a linearized system defining the relative dynamics of the two vehicles. This simulation was then validated by comparison to a truth model built in Goddard Space Flight Center's GMAT mission planning tool. The mission concept calls for only one of the two vehicles called the FollowerSat to carry propulsion, the second spacecraft called the LeaderSat will free fly in a natural orbit.

---

[*] Graduate Student, Mechanical and Aerospace Engineering, New Mexico State University, P.O. Box 30001, MSC 3-0 Las Cruces, NM 88003.
[†] Professor, Klipsch School of Electrical and Computer Engineering, New Mexico State University, P.O. Box 30001, MSC 3-0 Las Cruces, NM 88003.
[‡] Associate Branch Head, Navigation and Mission Design Branch, NASA Goddard Space Flight Center, Mail Code 595 Greenbelt, MD 20771.
[§] Sr. Scientist, Astroparticle Physics Laboratory, University Space Research Association, Mail Code 661 Greenbelt, MD 20771.
[**] Professor, Department of Mechanical Engineering, University of New Mexico, Mechanical Engineering Department 1 University of New Mexico MSC01 1150, Albuquerque, NM 87131.

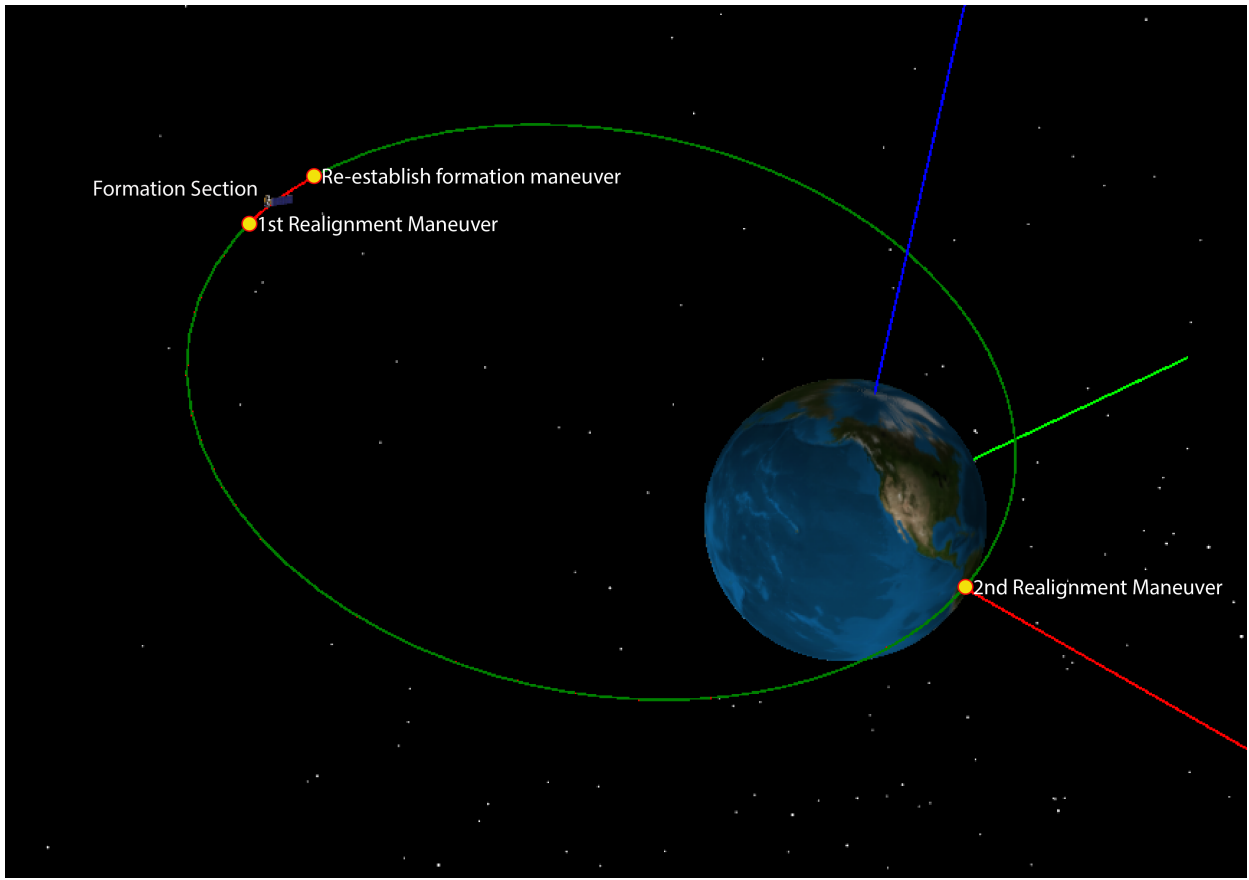

**Figure 1: Image shows an overview of the mission design concept for VTXO.**

**ASSUMPTIONS**

The simulation makes several assumptions in its present form. The model assumes that the leader spacecraft follows a purely Keplerian orbit around the Earth, with no inclusion of perturbations due to the Moon, Sun, atmospheric effects, electromagnetic effects, or solar radiation pressures. These assumptions are reasonable since when modeling relative dynamics, the primary concern is the difference between these forces interactions on each of the two vehicles. Given that the two vehicles remain in relatively close proximity to each other at all times the difference in the forces will be small.

**SIMULATION DESIGN**

The simulation is based on the following equations reformulated from a 2012 paper by Calhoun and Shah [1]. Calhoun, and Shah derive the following linearized equation.

$$\ddot{\overline{\delta_R}} = \Gamma_{GG}\overline{\delta_R} + \Gamma_{GG}\overline{R_f}^{ref} + \overline{u_R} \tag{1}$$

Where:

$[\Gamma_{GG}]$ - 3x3 matrix representing the differential gravitational forces between the two vehicles

$$[\Gamma_{GG}] = -\frac{\mu}{\left\|\overline{r_f^{ref}}\right\|^3}\left([I] - 3\widehat{r_f^{ref}}\left[\widehat{r_f^{ref}}\right]^T\right) \tag{2}$$

$\overline{r_f^{ref}}$ - Vector from the earth to the follower satellite

$\overline{R_f^{ref}}$ - Desired Vector between the two satellites (Telescope Axis)

$\overline{\delta_R}$ - Error between desired and actual follower satellite position

$\overline{u_R}$ – Propulsion, and orbital perturbations.

This was then converted into the following set of first order ODEs which can then be readily solved numerically.

$$\begin{Bmatrix}\ddot{\delta}_{Rx}\\\ddot{\delta}_{Ry}\\\ddot{\delta}_{Rz}\\\dot{\delta}_{Rx}\\\dot{\delta}_{Ry}\\\dot{\delta}_{Rz}\end{Bmatrix} = \begin{bmatrix}0 & 0 & 0 & & & \\0 & 0 & 0 & [\Gamma_{GG}] & & \\0 & 0 & 0 & & & \\1 & 0 & 0 & 0 & 0 & 0\\0 & 1 & 0 & 0 & 0 & 0\\0 & 0 & 1 & 0 & 0 & 0\end{bmatrix}\begin{Bmatrix}\dot{\delta}_{Rx}\\\dot{\delta}_{Ry}\\\dot{\delta}_{Rz}\\\delta_{Rx}\\\delta_{Ry}\\\delta_{Rz}\end{Bmatrix} + \begin{Bmatrix}[\Gamma_{GG}]\overline{R_f^{ref}}\\0\\0\\0\end{Bmatrix} + \begin{bmatrix}[K_d] & & & [K_p] & & \\0 & 0 & 0 & 0 & 0 & 0\\0 & 0 & 0 & 0 & 0 & 0\\0 & 0 & 0 & 0 & 0 & 0\end{bmatrix}\begin{Bmatrix}\dot{\delta}_{Rx}\\\dot{\delta}_{Ry}\\\dot{\delta}_{Rz}\\\delta_{Rx}\\\delta_{Ry}\\\delta_{Rz}\end{Bmatrix} \tag{3}$$

Where $\overline{u_R}$ contains a PD Control system.

$$\overline{u_R} = \begin{bmatrix}[K_d] & & & [K_p] & & \\0 & 0 & 0 & 0 & 0 & 0\\0 & 0 & 0 & 0 & 0 & 0\\0 & 0 & 0 & 0 & 0 & 0\end{bmatrix}\begin{Bmatrix}\dot{\delta}_{Rx}\\\dot{\delta}_{Ry}\\\dot{\delta}_{Rz}\\\delta_{Rx}\\\delta_{Ry}\\\delta_{Rz}\end{Bmatrix} \tag{4}$$

and

$[K_d]$ - Derivative gain matrix

$[K_p]$ - Proportional gain matrix

For all of the simulation runs presented in this report, the $K_d$, and $K_p$ matrix are both identity matrices. This is an area for potential improvement, however this seems to give adequate results for the purpose of this simulation. During a brief experiment, where the gains were adjusted to attempt to improve the simulation results, a substantial improvement in control tolerances was not achieved, nor was there a significant change in the results of the control system. The primary change that was observed was a substantial increase in the simulation run time as higher control gains forced the solver to run at smaller step sizes.

**Validation**

This simulation was validated using GMAT (General Mission Analysis Tool) as a truth model. GMAT is a mission design program written by GSFC (Goddard Space Flight Center), it contains detailed and well validated orbit models, and has been used by several successful missions. As such, it produces reliable orbit data to use as a truth model.

**Validation at 100m Initial Separation**

During validation the control gains in the simulation were set to zero, which models the effects of the two spacecraft drifting freely. The initial validation starts with $\overline{R_f}^{ref} = \begin{Bmatrix} 0.1 \\ 0 \\ 0 \end{Bmatrix} km$. The simulation is then run with a variety of orbit parameters. After four orbits (~40h), the difference between the position of the MATLAB simulation, and the truth model is around 2% of the distance between the leader and the follower spacecraft.

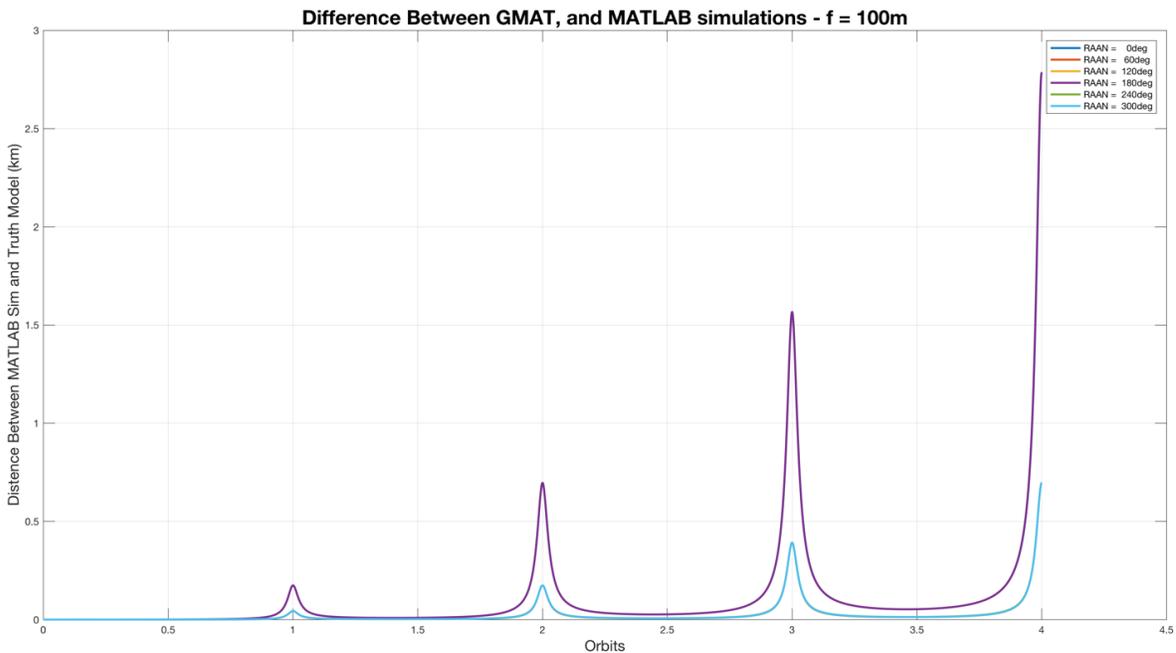

**Figure 2: Plot shows magnitude of the distance between the truth model, and my simulation over time for several sets of orbits with 100m initial separations.**

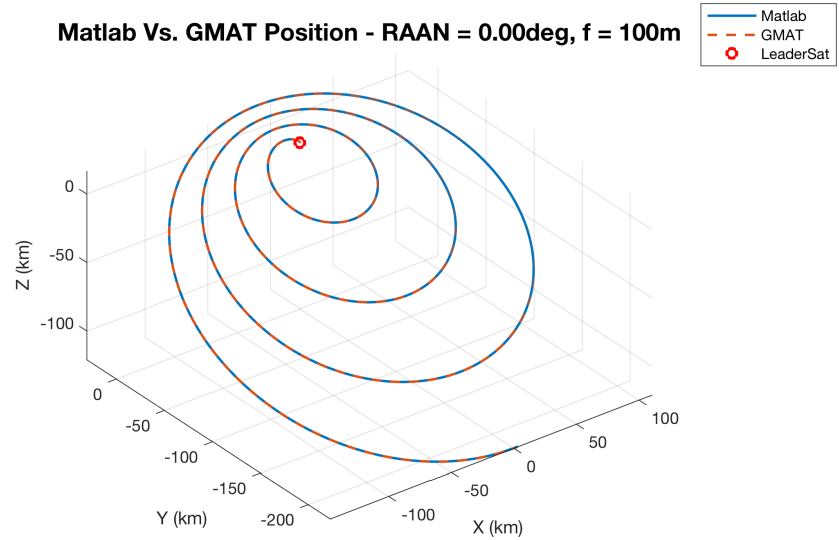

**Figure 3: Representative plot showing the paths relative to the LeaderSat of the FollowerSat for both the truth model, and the MATLAB simulation with 100m initial separations.**

**Validation at 1,000m Initial Separations**

The validation was then run with an initial condition of $\overline{R_f^{ref}} = \begin{Bmatrix} 0.1 \\ 0 \\ 0 \end{Bmatrix} km$, with these conditions after four orbits the error between the MATLAB simulation, and the GMAT truth model is close to 10% of the distance between the LeaderSat, and the FollowerSat for the worst case. However, by this point the Leader, and FollowerSat are over 2000km apart from each other, at this distance, is not surprising that the linearized equations used in the MATLAB simulation are breaking down. It is also worth noting that at this distance the assumption that the orbit perturbations have a minimal effect on the relative dynamics is no longer valid.

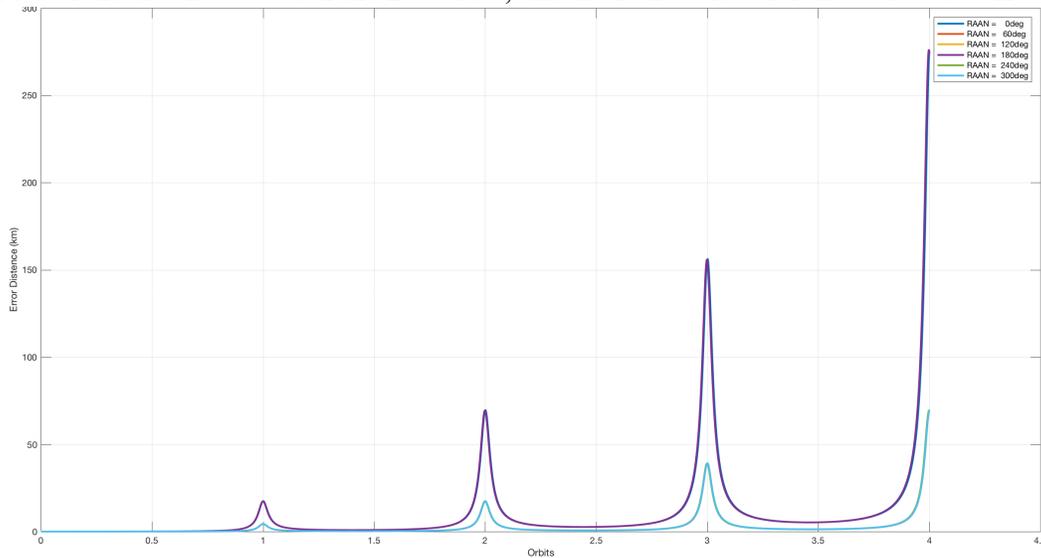

**Figure 4: Plot shows separations between MATLAB simulation, and GMAT truth model for 1000m initial separations over a range of orbits.**

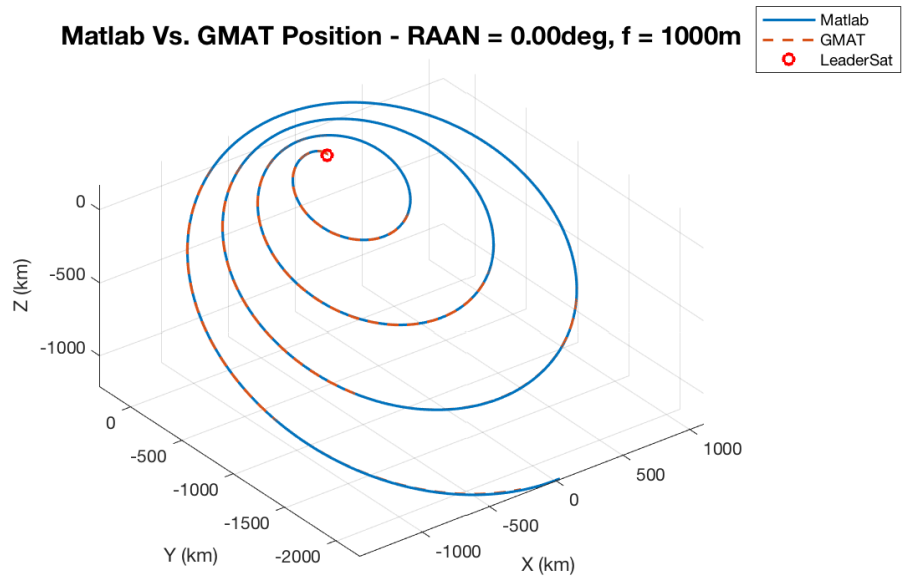

**Figure 5: Representative plot showing the paths relative to the LeaderSat of the FollowerSat for both the truth model, and the MATLAB simulation with 1000m initial separations.**

**Validation Results**

Given that in this model, the two spacecraft are continually moving apart, it is difficult to quantify a specific distance where this simulation breaks down from these tests, however it is clear that the model provides good results out to a few kilometers separation, and continues to provide usable results out to a few tens of kilometers separation. It is likely that the simulation could be used at even further distances, particularly if some of the orbital perturbation terms where included. However, a more rigorous method of validation should be used before this model is used for spacecraft separations exceeding a few tens of kilometers.

**ANALYSIS**

The simulation then had a controller implemented into it. By analyzing the amount of acceleration that was commanded by the controller, it is possible to examine the FollowerSat's fuel consumption under a variety of conditions.

**Variations in Telescope Pointing Axis**

The first analysis run was pointing the telescope in various pointing axes with the following orbital elements, which are consistent with a GTO (Geostationary Transfer Orbit).

Inclination = 0deg

RAAN = 0deg

Initial True Anomaly = 0deg

Argument of Perigee = 0deg

Perigee = 6778km (400km altitude)

Apogee = 42164km (35,786km altitude)

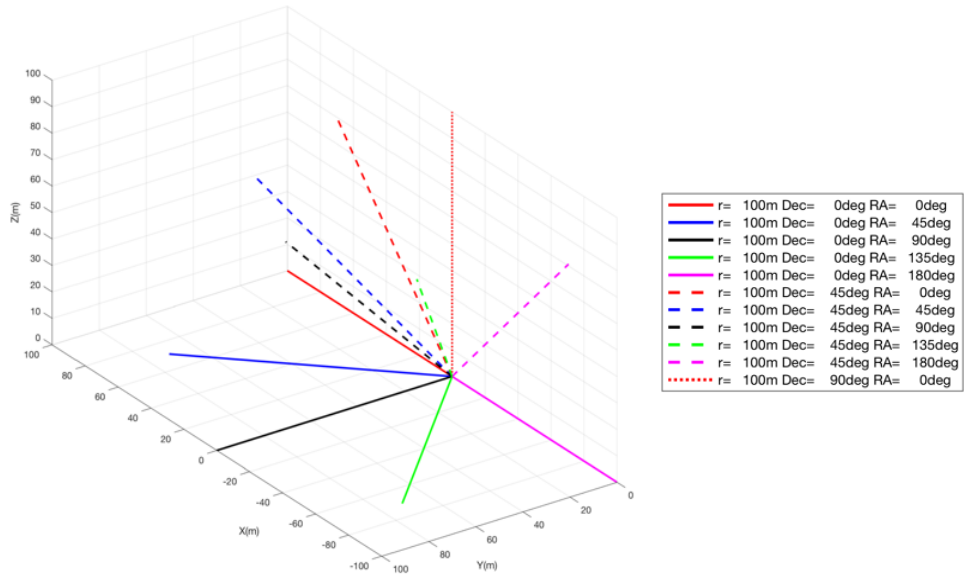

**Figure 6: Depiction: of the various modeled telescope axis. The negative x axis is aligned with the orbits apse line, and is pointing towards perigee. The line colors and types are constant across all plots.**

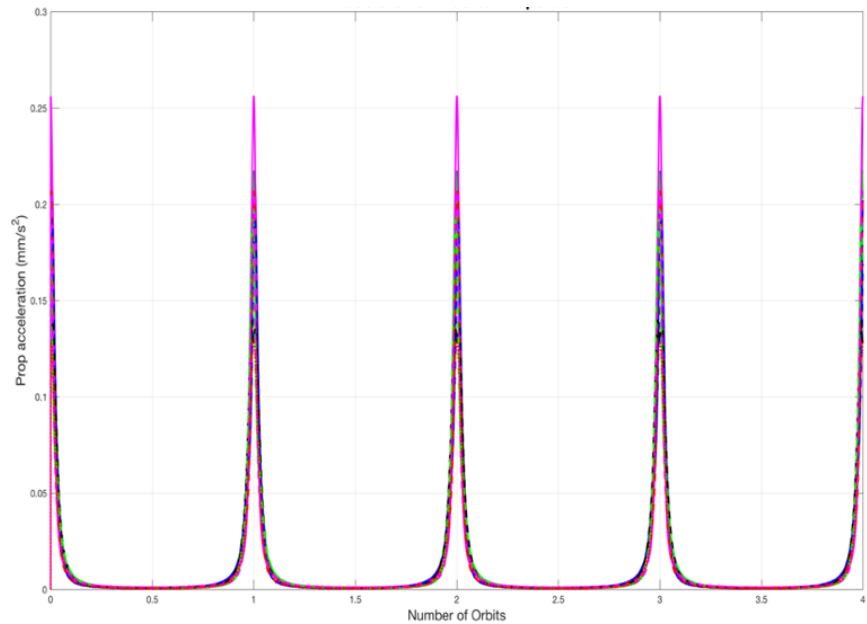

**Figure 7: Plot showing the acceleration generated by the propulsion system over four orbits**

Figure 7 shows the acceleration generated by the propulsion system over the duration of four orbits. The key item to note in this plot is that as anticipated the required propulsion (fuel consumption) is dramatically larger near perigee, and then drops off near apogee.

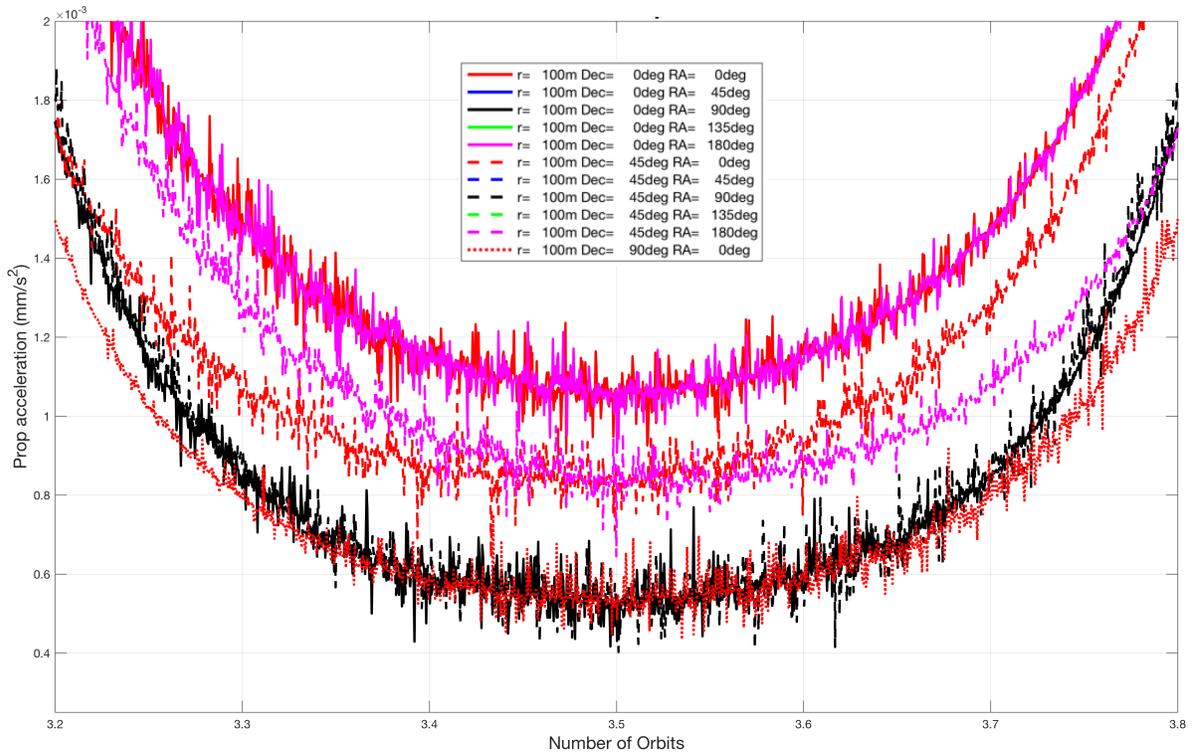

**Figure 8: Shows the acceleration generated by the propulsion system in order to produce the desired formation over a range of telescope axis.**

Figure 8 then zooms into the area near apogee for one of the orbits. The first important item to note is that the accelerations tend to group by the right ascension angle (note that at a declination of 90deg, the RA is undefined), with the curves that generate optimum performance being those that lie on the plane normal to the orbit's apse line. Another minor note to be made from this plot is that for several of the orientations, the curves do not center on apogee. This is not inherently problematic, but worth watching during mission design.

Once these results were generated, the results were re-run with more pointing directions. As Figure 9 shows, the trends continue when more pointing directions are considered. An unexpected result is that not all pointing directions have their minimum fuel consumption perfectly centered on apogee.

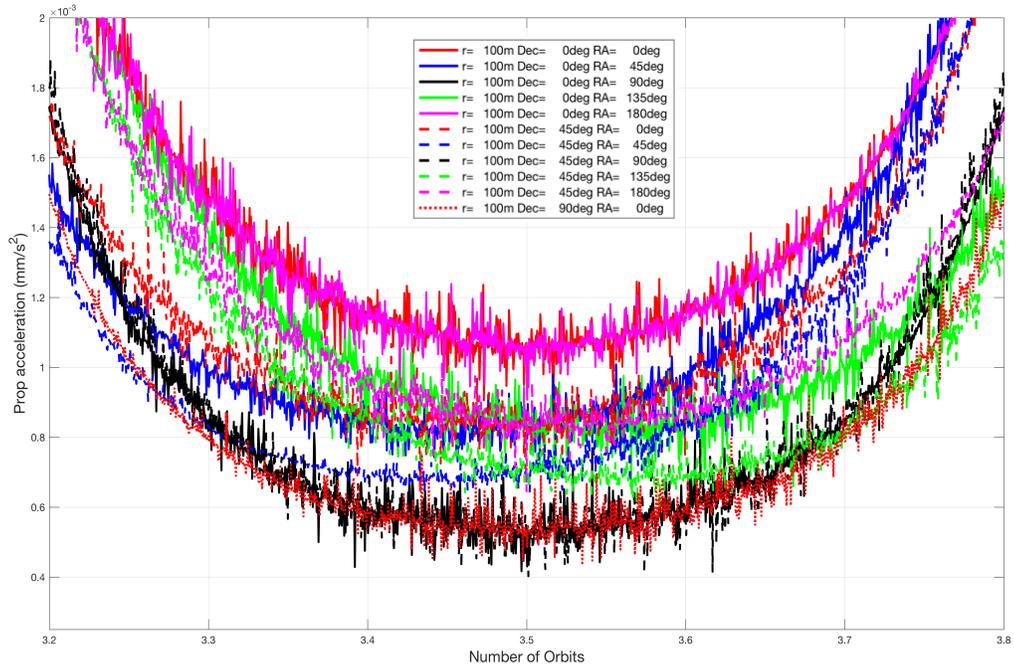

**Figure 9: Plot of acceleration generated by the propulsion system near apogee for a large number of telescope pointing conditions.**

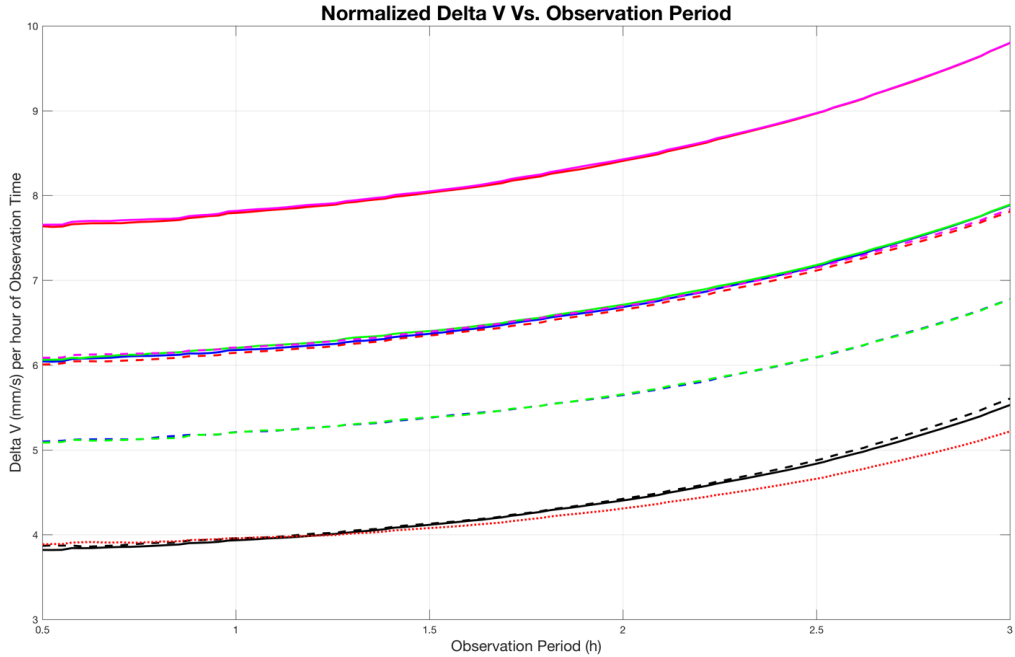

**Figure 10: Plot showing the ΔV required for a given observation time. The results have been normalized to a per hour basis. Note that this plot does not include the fuel needed to re-position the vehicles back into formation between observation periods.**

**ΔV Required to Maintain Formation**

The next analysis that was run was to calculate the ΔV required to maintain formation by integrating under the acceleration due to propulsion curves. By varying the time interval that is looked at, it becomes possible to ascertain the amount of ΔV required to maintain formation for a given interval. These results are then normalized to be in terms of per hour of observation, so as to understand the relationship between individual observation periods, and total observations over the mission. These results can be seen in Figure 10, as would be expected, the results show that the longer the observation period is, and in turn the further from apogee the formation is held, the larger the fuel consumption is per hour of observation. However, the curves are relatively flat out to 1h – 1.5h of observations, meaning observations can be made for at least this long with minimal penalties in overall observation time. An important note is that this plot only shows the ΔV consumed during the formation period, and does not include the ΔV required to maneuver the vehicles back into formation between observation periods.

**Variations in Focal Length**

When looking at variations in the telescope focal length (how far apart the two spacecraft are), there appears to be a nearly linear relationship between the focal length, and ΔV. The results show a fully linear relationship, which is likely the result of the linearized equations being used for this model. However, given that the simulation validation shows excellent results at the longest focal length being considered (1km), the linearized results should be very close to accurate in the range of interest (100m – 1,000m).

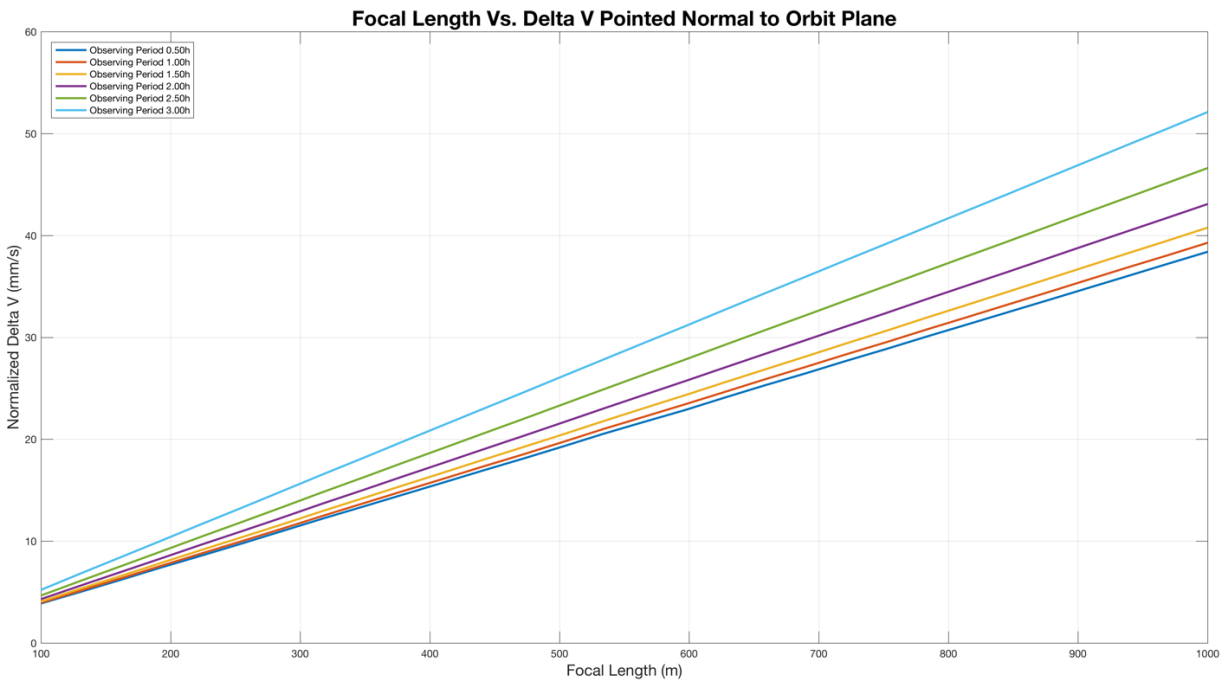

**Figure 11: Plot of telescope focal length vs. the ΔV consumption on a per hour basis. Plot has vehicles in the optimum pointing direction of normal to the plane of the orbit.**

**Mission Lifetime**

By making a few assumptions, it is possible to get a very rough estimation of the lifetime capabilities of a mission such as VTXO. By estimating that VTXO will have 100m/s of total ΔV capability. Then make a conservative estimate that VTXO will use on average 100mm/s of ΔV per hour of observations, approximately an order of magnitude greater than this analysis estimates. Using these numbers, it can be estimated that VTXO will be capable of around 1,000h of total observation time.

**CONCLUSIONS**

There are a few major conclusions which can be drawn from this work about the VTXO mission. The first major conclusion is that the optimum pointing direction for the telescope is with the telescope axis pointed normal to the plane of the orbit. Near optimum performance can be achieved with the telescope axis lying anywhere on the plane normal to the orbit's apse line.

The second major conclusion is that there is a minimal penalty to pay in terms of fuel for making continuous observations in excess of an hour. Additionally, observation periods of several hours (2h – 4h) are viable with relatively small penalties (10% - 30%) in terms of fuel consumption.

Finally, over the course of the mission, well over 1,000h of observation time should be readily achievable.